\documentclass{optica-article}

\journal{opticajournal} % for journals or Optica Open

\articletype{}

\usepackage{lineno}
\usepackage{hyperref}
\usepackage{physics}
\usepackage{comment}
\usepackage{pdfpages}
\nolinenumbers

\begin{document}

% what is the story:  integrated all together at once, move some background into setup discussion, everything is matched. 

\title{High-performance source of indistinguishable polarization-entangled photons with a local oscillator reference for quantum networking}

\author{\fontsize{9pt}{9pt}\selectfont Michael Grayson,\authormark{1,*}, Shawn Meyer \authormark{2}, Daniel Sorensen\authormark{1,2}, Abigail Gookin \authormark{1,2}, Markus Allgaier \authormark{3,4}, Nicholas V. Nardelli \authormark{1}, Tara M. Fortier \authormark{1}, Dileep V. Reddy \authormark{1,4},  Martin J. Stevens\authormark{1}, Michael D. Mazurek \authormark{1,4}, Juliet T. Gopinath \authormark{2,4}, and L. Krister Shalm \authormark{1,2}}

\address{\fontsize{8pt}{8pt}\selectfont \textit{\authormark{1} National Institute of Standards and Technology, Boulder, CO 80305\\
\authormark{2} Electrical, Computer and Energy Engineering Department, University of Colorado, Boulder, CO 80309\\
\authormark{3} Department of Physics and Astrophysics, University of North Dakota, Grand Forks, ND 58202\\
\authormark{4} Department of Physics, University of Colorado, Boulder, CO 80309\\
}}

\email{\authormark{*} michael.grayson@colorado.edu} %% email address is required; see note below about the corresponding author designation

% use {asbstract*} to suppress the copyright line. Copyright information will be added in production

\begin{abstract*} 

Optical quantum networking protocols impose stringent requirements on the states produced by sources of entanglement. We demonstrate a free-space, compact, source of indistinguishable pairs of polarization entangled photons, with an integrated local oscillator reference as a significant step towards this goal. This source achieves $(99.11 \pm 0.01) \%$ polarization entanglement visibility, $(96.3 \pm 0.6) \%$ successive-photon Hong-Ou-Mandel interference visibility, $(68.0 \pm 0.1$) \% heralded efficiency as detected, and $(88.6 \pm 0.2) \%$ interference visibility with a local oscillator. This simultaneous achievement of state-of-the-art metrics demonstrates an adaptable platform for quantum networking.

\end{abstract*}

\section{Introduction}
Optical quantum networking platforms distribute quantum information using photons, which naturally provide a high-speed, low decoherence quantum channel that is compatible with existing fiber communication infrastructure\cite{wei_towards_2022, lafler_optimal_2022}. Applications of optical quantum networking include secure communication \cite{shalm_device-independent_2021,yin_entanglement-based_2020}, distributed sensing \cite{wang_astronomical_2025, guo_distributed_2020}, and computing \cite{main_distributed_2025}, each of which has specific protocols developed for near term implementation at telecom wavelengths that utilize the existing low loss fiber optic classical network. Although these protocols share common requirements on photon sources, each imposes distinct stringent demands--such as indistinguishability of photons \cite{lafler_optimal_2022,pickston_optimised_2021}, entanglement quality \cite{shalm_device-independent_2021, buhrman_quantum_2016-1}, or phase stability \cite{alwehaibi_tractable_2025}--making it difficult to develop a single source that is appropriate for all protocols. Here we present a flexible telecom-wavelength entangled-photon platform that, for the first time, combines multiple capabilities co-designed to support a wide range of quantum networking applications. This platform is able to generate indistinguishable photon pairs needed for multi-source interference (required in entanglement swapping), entangled polarization photon pairs with high fidelity, high efficiency heralded single photons, and a spectral-temporal mode matched local oscillator (LO). Additionally, this entire system can be phase stabilized to be compatible with emerging path-entangled networking protocols. Together, these capabilities cover the requirements over a wide range of quantum networking protocols for telecom optical quantum networking.

% Krister %

Telecom wavelength photons are ideal for carrying out a large set of quantum communication and sensing protocols \cite{wei_towards_2022, shalm_device-independent_2021, guo_distributed_2020}. Protocols such as entanglement-based quantum key distribution (QKD) or short-range device-independent random number generators (DI-RNG) require high-quality two-photon entanglement \cite{kavuri_traceable_2025,qi_experimental_2007}. These entangled-qubit sources encode the entanglement in degrees of freedom such as polarization or time bins. Other protocols, such as quantum-enhanced very long baseline interferometry \cite{gottesman_longer-baseline_2012} or longer distance device-independent communications protocols \cite{alwehaibi_tractable_2025}, use single-photon entanglement (path-entanglement) and an LO to perform displacement measurements \cite{paris_displacement_1996}. This can require a pure source of heralded single photons that can be interfered with an LO. Additionally, all quantum optical protocols face loss that is exponential in transmission distance \cite{dhara_entangling_2023}. Depending on the protocol, there is a distance at which a quantum repeater is required to overcome the exponential loss \cite{dhara_entangling_2023}. This involves interfering photons from different sources to "swap" the entanglement between distant nodes. Entanglement swapping requires photons from different sources to be indistinguishable from each other \cite{pickston_optimised_2021, graffitti_independent_2018}. 

The current approach to building optical quantum networks is to develop application-specific entangled-photon sources \cite{shalm_strong_2015, pickston_optimised_2021, qi_experimental_2007, xu_characterization_2023}. Therefore, a quantum network supporting multiple protocols may require many different sources, which can become inefficient. There is a need for a flexible approach to building optical entangled sources that can be scaled to support multiple different protocols and qubit encodings. Building such a source faces multiple challenges. A standard approach for building sources is to use an optical nonlinear process, such as type-II spontaneous parametric down-conversion (SPDC), which probabilistically generates photon pairs. This can be combined with a system that erases which-path information to produce polarization entanglement \cite{shalm_strong_2015, kwiat_ultrabright_1999, steinlechner_high-brightness_2012, meyer-scott_high-performance_2018, jabir_robust_2017, gerrits_generation_2011}. %extend this setence and cite two mode ent sources

A major challenge with SPDC is that the photons generated by such a process may be highly-entangled in their spectral degree of freedom due to energy conservation \cite{pickston_optimised_2021, graffitti_independent_2018, lafler_optimal_2022, dixon_spectral_2013}. This unwanted entanglement leads to a reduction in the purity/indistinguishability of the photons produced, rendering them unsuitable for applications requiring multi-photon interference\cite{pickston_optimised_2021,lafler_optimal_2022}. Careful control of the pump temporal-spectral and spatial properties, as well as engineering the phasematching of the nonlinear process, can lead to photon pairs that are spectrally separable without filtering \cite{pickston_optimised_2021, harder_optimized_2013,lafler_optimal_2022, jeronimo-moreno_control_2009, sanchez-lozano_relationship_2012, meyer-scott_limits_2017}. In some applications, the photons generated interfere with an LO (a coherent state). This requires the LO to occupy the same spectral-temporal mode as the photons produced by the SPDC source \cite{gottesman_longer-baseline_2012, allgaier_highly_2017, guo_distributed_2020}. Many of these applications are sensitive to wavelength scale path-length fluctuations that can introduce relative phase shifts \cite{alwehaibi_tractable_2025, wang_astronomical_2025}. For advanced protocols, this can require remote LOs to be phase locked to each other over a network. To help with this locking, it is desirable to have the LO and the down-conversion photons be derived from the same laser. However, the lasers used to pump the nonlinear process have twice the energy of the generated photons in SPDC and are therefore not indistinguishable. Finally, it is desirable to have both a high rate of photons generated and to couple those photons as efficiently as possible into fiber. However, there is a fundamental trade-off between rates and collection efficiency with bulk SPDC sources \cite{bennink_optimal_2010}. High rate sources require tight focusing conditions for the pump beam as well as the down-conversion modes, while high-efficiency coupling to single-mode fibers requires looser focusing conditions \cite{bennink_optimal_2010}. Although progress has been made tackling each of these problems individually, to the best of our knowledge, no single source is capable of supporting all of these capabilities. Here we present a first generation telecom quantum source platform that can produce indistinguishable photons, supports high-quality polarization-entangled photons, is capable of low-loss fiber coupling or high entanglement-generation rates, and has an optional LO that can be used for displacement measurements. 

To construct a platform that supports all of these capabilities, we first designed a down-conversion process to produce spectrally separable photons. We chose commercially available periodically poled potassium-titanyl phosphate (ppKTP) as the SPDC medium which at 1550 nm supports a type II down-conversion process whose group velocity conditions produce a symmetric distribution of emitted signal and idler energies, known as the joint spectral intensity (JSI) \cite{dixon_spectral_2013,pickston_optimised_2021,graffitti_independent_2018}. To make our SPDC photons more resistant to fiber induced dispersion, we chose the maximum length commercially available crystal (30 mm) to produce the narrowest bandwidth (approximately 2 nm). We designed the crystal to remove residual spectral correlations -- from the finite length of the crystal -- by apodizing the effective nonlinearity across the crystal through duty cycle modulation \cite{dixon_spectral_2013, pickston_optimised_2021, graffitti_independent_2018}.

Producing indistinguishable photons requires the temporal and spectral profile of the pump to exactly match the photons generated in the down-conversion process \cite{pickston_optimised_2021,graffitti_independent_2018,dixon_spectral_2013, sanchez-lozano_relationship_2012}. To the best of our knowledge, no such commercial source in the telecom exists. Therefore, we built a custom pump laser that can be phase stabilized and provides both an LO and 775 nm pump which can be tuned to match our down-conversion process. For the seed we chose a 1550 nm modelocked laser, which can be made frequency stable to one part in $10^{18}$ \cite{nardelli_10_2022}. This seed laser produces an approximately 11 nm bandwidth, which we spectrally filter to match the approximately 2 nm bandwidth of the down-conversion process. To avoid the difficulty of building a separately tuned pump laser, we engineer a nonlinear optical process which converts the LO to a 775 nm pump while maintaining the relative spectral-temporal properties of the LO. This is done through chirped pulse amplification followed by second harmonic generation -- both engineered to leave the pulse transform-limited -- resulting in the matching of both the LO and the pump to the SPDC photons without an additional tuning process.  

Type II down-conversion can be used to produce polarization entanglement or heralded single photons. Producing polarization entanglement requires the erasure of which-path information between two modes of down-conversion, while a heralded single photon requires separation of two photons from a single down-conversion process \cite{shalm_device-independent_2021}. To enable both in our platform, we engineered a beam-displacer-based Mach-Zehnder interferometer (MZI) which can either produce polarization entanglement or heralded single photons by tuning a single waveplate. This entire system was engineered to be compact and passively aligned using a machined aluminum mount for the polarization optics. Finally, we designed an optical system to integrate with all of these components and provide a modest trade off in heralding efficiency versus rate. We target a $90\%$ maximum heralding efficiency, sufficient for device independent (DI) protocols, while also providing a high rate of SPDC photons. We used a combination of gradient refractive index (GRIN) lenses and plano convex lenses to achieve these waists while providing low losses.

While portions of this platform were co-designed (the SPDC crystal and filtered pump), many of the other components are modular. The pump filter is tunable to accommodate a range of bandwidths for alternative crystal designs. The MZI can be tuned to provide a variety of polarization entangled states, which could be further converted into other types of two-mode entanglement. Additionally, the MZI could be replaced with an entirely different module without affecting the indistinguishability of the photons. The pump and collection lenses are also largely independent from the indistinguishability, and could be tuned to provide higher rates or higher heralding efficiency.

\section{Experimental Setup}
\begin{figure}[htbp]
\centering\includegraphics[width=14cm]{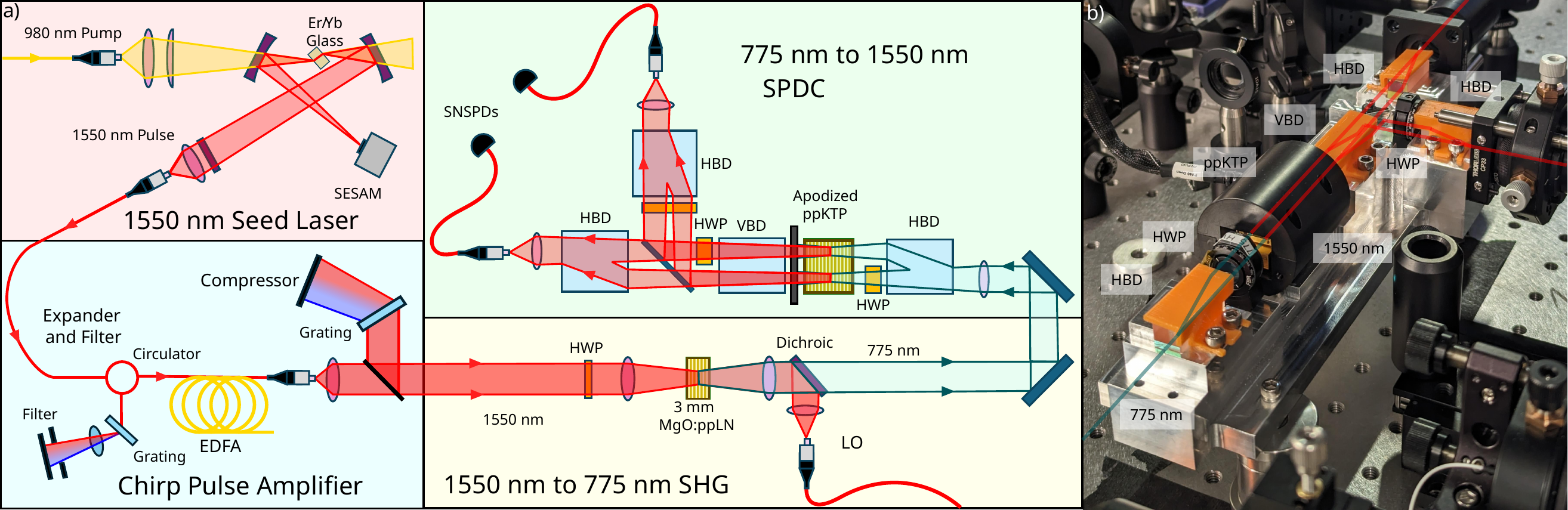}
\caption{a) A simplified depiction of our SPDC source. We use a 980 nm pump laser to excite Er:Yb doped glass in a 100 MHz cavity to produce a broad spectrum seed. This acts as both a stable seed for the down-conversion and the LO reference. We filter and amplify this seed using Martinez-style compressors and an erbium doped fiber amplifier (EDFA) \cite{backus_high_1998,chauhan_single-diffraction-grating_2010, martinez_grating_1986,zhong_investigation_2020, ruiz-de-la-cruz_multi-pass_2005}. This allows us to match the seed spectrum to the apodized down-conversion crystal and produce a high-power pulse. We then up-convert this 1550 nm light to 775 nm using a commercial magnesium-oxide-doped, periodically poled lithium niobate (MgO:ppLN) crystal. We down-convert this light using apodized periodically poled potassium titanyl phosphate (ppKTP) in a beam displacer Mach-Zehnder interferometer (MZI). This produces indistinguishable polarization-entangled photons that we fiber couple with high efficiency and detect with superconducting nanowire single-photon detectors (SNSPDs). b) An image of the monolithic aluminum mount used to passively align all polarization optics in our down-conversion system. The beam displacers are held in by orange 3D printed tabs. The waveplates are mounted using compact rotation mounts. Finally, the crystal is mounted in a black crystal oven. }
\label{fig:layout}
\end{figure}

To meet the demands of quantum networking, we integrate four primary subsystems, each co-designed. These four systems are outlined in Fig. \ref{fig:layout}. We use a 980 nm continuous wave (CW) diode laser to pump a 100 MHz repetition rate mode-locked laser to generate 1550 nm femtosecond pulsed light \cite{nardelli_optical_2022}. This laser is both a stable seed for the photon pairs and the LO that is matched to the down-conversion. We then filter and amplify the seed using chirped pulse amplification (CPA) \cite{backus_high_1998,chauhan_single-diffraction-grating_2010, martinez_grating_1986,zhong_investigation_2020, ruiz-de-la-cruz_multi-pass_2005}. This allows us to tune the spectrum of the seed while removing the temporal correlations in the pump laser. We then convert the 1550 nm seed to 775 nm using bulk second harmonic generation (SHG) and collect the unconverted 1550 nm light as the LO \cite{boyd_chapter_2008}. Finally, we generate pairs of photons at 1550 nm through type-II spontaneous parametric down-conversion and collect them efficiently into fiber \cite{shalm_device-independent_2021, shalm_strong_2015, pickston_optimised_2021}. The crystal generating the type-II process is apodized to remove any spectral correlations in the produced photons \cite{dixon_spectral_2013,graffitti_independent_2018,pickston_optimised_2021}.

Having low cost and stable components that can be used to build a quantum network is imperative for a scalable single-photon source. Our seed laser leverages mature telecommunication technologies such as 980 nm telecom pump diodes, which have an order of magnitude longer operating lifetimes than the pumps of titanium-sapphire lasers\cite{nardelli_optical_2023, casteele_high_2008}. Additionally, this laser has demonstrated an active fractional frequency stability of one part in $(10^{-18})$ \cite{nardelli_optical_2023}. However, it is not carrier-phase locked in our experiment. We use a 980 nm continuous-wave diode laser to pump an Er/Yb:glass in a 100 MHz repetition rate, passively mode-locked 1550 nm cavity. This produces a soliton with a $(0.23 \pm 0.03)$ ps duration with an $(11.1 \pm 0.1)$ nm full width half maximum (FWHM) bandwidth at $(29.53 \pm 0.02)$ mW average power. We measure the temporal width using an autocorrelator assuming a $\text{sech}^{2}$ profile. This pulse is too broadband for kilometer-scale telecommunications and too low power for SPDC, so it is filtered and amplified through chirped pulse amplification.

The purity of generated photons depends significantly upon the spectral-temporal correlations of the pump pulse \cite{graffitti_independent_2018,pickston_optimised_2021, sanchez-lozano_relationship_2012}. We achieve low spectral-temporal correlations using a CPA scheme specifically matched to our down-conversion crystal. The pulse is expanded in time using a multi-pass Martinez expander \cite{martinez_grating_1986, ruiz-de-la-cruz_multi-pass_2005}. Light enters and exits the expander through a fiber circulator, which separates the input pulse from the expanded and filtered output. The Martinez expander uses four passes through a $1000$ grooves/mm grating with 52 degree incidence and 20 cm lens, which is de-focused by $(12.7 \pm 0.4 )$ cm, providing a total of $(-14.4 \pm 0.6)$  ps$^2$ of group delay dispersion. The pulse is filtered using a tunable slit aperture while its frequencies are spatially dispersed by the grating. The expander produces a $(5.42 \pm 0.03)$ nm FWHM bandwidth pulse with a total efficiency of $(8.9 \pm 0.2) \%$. This is then amplified using a commercial erbium-doped fiber amplifier (EDFA) to $(551.5 \pm 0.2)$ mW. We then recompress the pulse using a four-pass, free-space Treacy compressor that has a tunable delay \cite{chauhan_single-diffraction-grating_2010, zhong_investigation_2020}. This produces a $(1.37 \pm 0.02)$ ps FWHM duration pulse with a $(3.16 \pm 0.03)$ nm FWHM bandwidth at $(362.8 \pm 0.5 )$ mW average power. The pulse duration was measured using an autocorellator assuming a Gaussian profile. Finally, we focus this pulse into a MgO:ppLN crystal that is  3 mm long with a 19.2 $\upmu$m poling period and is heated to $70^{\circ}$C. This crystal is designed to have a bandwidth large enough to leave the pump bandwidth unaffected. This SHG process produces a $(0.608 \pm 0.006)$ nm FWHM bandwidth 775 nm pulse at an average power of $(67 \pm 2)$ mW. Uncertainties in power were estimated by collecting power statistics for 1 minute. The uncertainty in bandwidth is estimated by the 95$\%$ confidence interval of the FWHM from a Gaussian fit of the relevant spectra. 

We use a beam displacer-based type‑II SPDC process in a periodically poled potassium titanyl phosphate (ppKTP) crystal to produce pairs of photons \cite{shalm_device-independent_2021, shalm_strong_2015}. The design has been shown to produce stable, high-quality polarization entanglement, ideal for quantum networking protocols. A single pump spatial mode enters the Mach‑Zehnder interferometer (MZI), while two spatially separated down-conversion spatial modes exit \cite{shalm_device-independent_2021}. We mount the polarization optics of our source on a CNC aluminum mount to passively align the polarization states and remove degrees of freedom, as shown in Fig. \ref{fig:layout}b. We focus both the pump and an alignment beam through the Mach-Zehnder paths using an achromatic doublet with a focal length of 15 cm. %removed machzhender description
Spectral information can be used to distinguish photons, which will affect the purity of quantum networking operations. To prevent this, we use a previously demonstrated technique known as apodization \cite{dixon_spectral_2013,graffitti_independent_2018,pickston_optimised_2021}. Without apodization, the periodically poled crystal applies a rectangular window on the spatial nonlinearity of the crystal, which induces a sinc dependence in the frequency domain \cite{dosseva_shaping_2016}. The side lobes of this sinc function will contain entangling spectral information between the signal and idler photons \cite{dosseva_shaping_2016, branczyk_engineered_2011, dixon_spectral_2013, pickston_optimised_2021}. We generate a Gaussian phase-matching function using a Gaussian apodization of the poling duty cycle. 

Our apodized ppKTP crystal is 27.5 mm long with a poling period of 46.5 $\mu$m. The profile has a FWHM apodization length of 14.6 mm. For a 0.6 nm pump FWHM bandwidth, the crystal is designed to produce signal and idler photons with bandwidths of 1.69 nm and 1.78 nm, respectively, with a Schmidt number of 1.0016. By generating a Schmidt number close to one, we ensure that the generated photons are indistinguishable and spectrally pure for high fidelity networking operations. The design simulation can be viewed at \href{https://app.spdcalc.org/#/?panels=W3sidHlwZSI6ImpvaW50LXNwZWN0cnVtIiwic2V0dGluZ3MiOnsiZW5hYmxlTG9nU2NhbGUiOmZhbHNlLCJoaWdobGlnaHRaZXJvyRZheGlzVMZVV2F2ZWxlbmd0aCIsImF1dG9VcGRhdGUiOnRydWV9fSzJflBhbmVsTG9hZGVyznt9fV0%3D&cfg=eyJhdXRvQ2FsY1RoZXRhIjp0cnVlLMkVUGVyaW9kaWNQb2xpbmfQHkludGVncmF0aW9uTGltaXRzyCFzcGRDb25maWciOnsiY3J5c3RhbCI6IktUUCIsInBtX3R5cGUiOiJUeXBlMl9lX2VvIizIJ1905gCIOTDKE3BoaSI6yxBsZW5ndGgiOjI3NTDLF3RlbXBlcmF0dXJlIjoyxBlvdW50ZXJfcHJvcGFn5QCoIjpmYWxzZSwiZmliZXJfY291cOwA2nB1bXBfd2F2ZchjNzc1xxZiYW5kd2lkxBUwLjbJK2lzdCI6NzDHEHNwZWN0cnVtX3RocmVzaG9sZMQuMDHHH3Bvd2VyIjoxLCJzaWduYWzNazE1NTDJGecBBsoR6AEExw%2FrAI7MTuUAjjHkARXME19wb3Np5gD%2BLTc5MjkuMjU2MTU2NTUwOTU2LCJpZGxlctMqNTcxLjgyMzM4NjUwMjU45AD05wIPX3DlAhBfZW5hYmxlZOkBPsYWxiYiOjQ2LjUyMDMyODUwMDYyMzM0LCJhcG9kaXrlAYrQPcwb5wIgR2F1c3NpYW4izh5md2htIjoxNDbkAOjMGXBhcmHEGs8Wb2lu5AKbW10sImRlZmbFIWnnArtvcuQCo21ldGjkALIiU2ltcHNvxHF0b2xlcmFuY2XlAVAwxGttYXhfZGVw5QF0xhJkZWdyZWUiOjTEDGl2cyI6NTB9fcpeaW9u6gMIbHNfbWlu5AHoNDYuODIsxRFheMQRNTMuxBBp0CFpziFzaXrmAI99fQ%3D%3D} {app.spdcalc.org} \cite{shalm_spdcalc_2022}. The targeted waist radius in the crystal is 100 $\mu$m for the pump beam and around 70 $\mu$m for the collection arms (signal/idler). We choose these sizes to provide a balance between optimal heralding efficiency and optimal pair generation rate \cite{bennink_optimal_2010}. This limits the maximum symmetric heralding efficiency to $90\%$, but enhances the pair generation rate relative to larger waists. We efficiently collect the light into fiber using two identical sets of collection optics. 

We collect the light using 15 cm lenses into a collimated beam, which is then focused into a commercial GRIN fiber coupler using a 40 cm lens. We fold both lens systems using $99.99\%$ reflectivity dielectric folding mirrors. We align the entire system with a 1550 nm alignment beam that is made collinear with the 775 nm pump beam using a dichroic mirror. These systems are designed to provide a high collection efficiency to enable secure quantum networking protocols. After collecting the photons into fiber, we efficiently detect them using in-house fabricated superconducting nanowire single-photon detectors (SNSPDs). The SNSPDs include a Bragg reflector and anti-reflective (AR) coating optimized for high efficiency detection at 1550 nm \cite{reddy_superconducting_2020}. These detectors are large in area and relatively slow, with a high jitter of ($223 \pm 2$) ps FWHM and a dead time of $(309 \pm 6)$ ns. The jitter reduces our wavelength resolution during JSI measurements to approximately 0.1 nm. We measured the system detection efficiency (SDE) of the four detectors to be $(93.4 \pm 0.5) \%$, $(96.3 \pm 0.5) \%$, $(92.8 \pm 0.5) \%$, $(93.4 \pm 0.5) \%$ \cite{reddy_broadband_2022}.

\section{Results}
We characterize this source by measuring heralded efficiency, joint spectral intensity (JSI), and Hong-Ou-Mandel (HOM) interference. We also measure second-order correlation, which is included in the supplement. We optimize each of these metrics via a feedback process to produce the high indistinguishability. We optimize the JSI using an adjustable slit in the Martinez stretcher that acts as the spectral filter on the pump. We optimize the HOM interference using the compressor delay. These settings deviated from the transform-limited case when optimized for purity, likely to due extraneous nonlinearities in both the CPA and SHG systems. The procedure first involves alignment of the system, then optimization of the herald efficiency, maximization of the JSI purity, and finally maximization of the successive-photon HOM visibility.
\subsection{Heralding and Rates}
We measure the heralded efficiency ($H$) using coincidences ($C$), and singles ($S$), calculating the symmetric heralded efficiency as 
\begin{equation} \label{eq:hera}
    H = \frac{C}{\sqrt{S_s S_i}},
\end{equation}
where $S_s$ is the singles count in the signal mode and $S_i$ is the singles count in the idler mode \cite{shalm_strong_2015, meyer-scott_limits_2017}. We monitor this value while iteratively maximizing singles in one arm and coincidences in the other. After this process, we obtain a $(68.0 \pm 0.1) \%$ symmetric heralded efficiency as detected with $128,600 \pm 400$ pairs at $(62.8 \pm 0.2)$ mW pump power incident on the SPDC crystal, integrated for 1 second. This corresponds to $2050 \pm 50$  pairs/(s mW). This efficiency is the heralded efficiency as measured with click events, the total heralded efficiency as detected, without correction for background counts, noise, or loss \cite{bienfang_single-photon_2023, migdall_absolute_1995}. We estimate the uncertainty assuming Poisson counting statistics. Having an efficiency well above $66.7\%$ is imperative for quantum networking applications with a large quantum advantage, which we approach in this experiment \cite{buhrman_quantum_2016-1}. Further efforts to increase the efficiency of the source such as more efficient detectors, and low loss splices will be required.

\subsection{Polarization Entanglement}
Once we obtain high efficiency, we use the beam displacer interferometer to generate polarization entanglement \cite{shalm_device-independent_2021 ,shalm_spdcalc_2022}. We do this by pumping two spatially separate modes of the nonlinear crystal and erasing the which-path information between photons generated in either mode \cite{shalm_device-independent_2021, shalm_strong_2015 , zhang_spontaneous_2021}. This process results in an entangled polarization state given by Eq. \ref{eq:state}
\begin{equation} \label{eq:state}
        \ket{\psi} = \frac{1}{\sqrt{2}}(\ket{HV} + e^{i \phi}\ket{VH}).
\end{equation}
where the relative phase $\phi$ depends on the MZI interferometer \cite{shalm_device-independent_2021, shalm_strong_2015 , zhang_spontaneous_2021}.

\begin{figure}[htbp]
\centering\includegraphics[width=12cm]{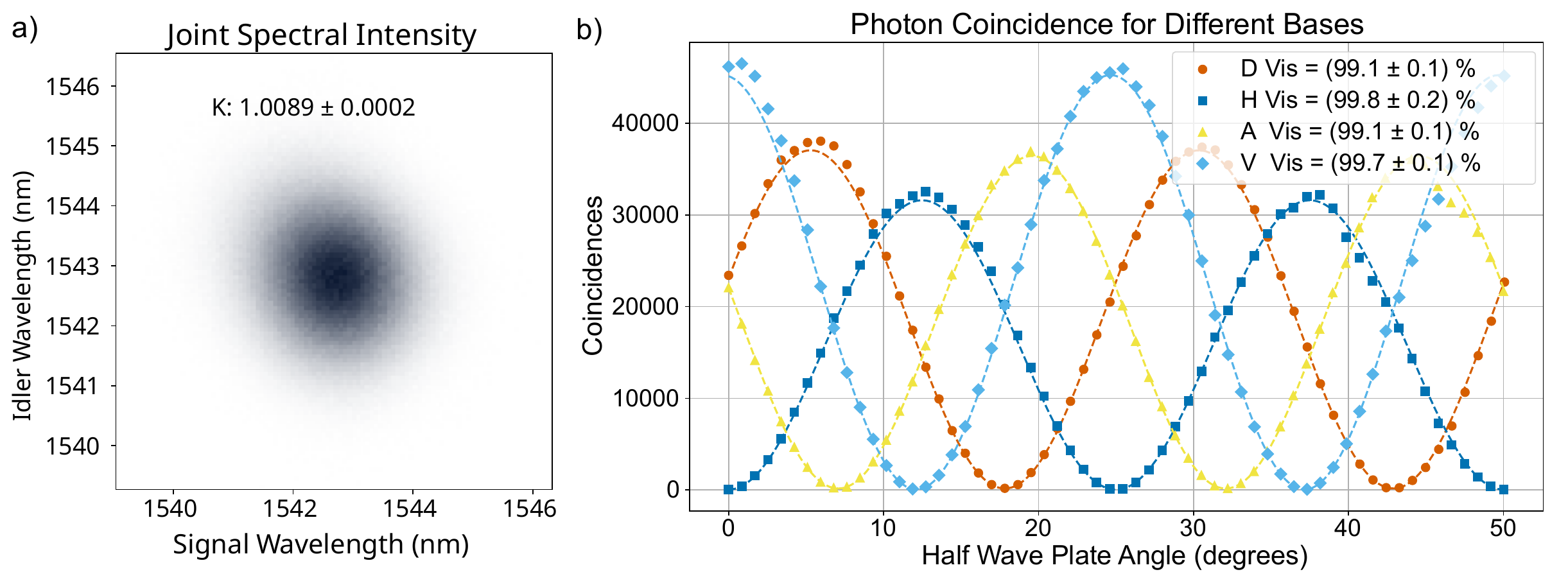}
\caption{a) Measured joint spectral intensity (JSI) of the photon pairs, obtained using time of flight spectroscopy. The JSI depicts the correlations in emission wavelengths in the signal and idler of generated pairs. We tune the crystal temperature and angle for degenerate photon pairs and low Schmidt number of $1.0089 \pm 0.0002$, calculated using Eq. \ref{eq:K}. This implies spectrally separable photons. b) Measured coincidences when we rotate the basis of the second polarization analyzer. The first polarization analyzer is set to the A,D,V,H basis. The change in absolute intensity for different bases is due to the polarization sensitivity of the detectors.}
\label{fig:pol_ent}
\end{figure}

Using two polarizers and motorized half waveplates on either collection arm, we measure the polarization state. We use one polarizer to change the basis while the other is rotated through a series of angles to measure the interference in that basis. We use a HWP on the pump beam to equalize the power in either state and tune the phase of the modes relative to each other by tilting a glass plate in one mode of the Mach-Zehnder interferometer. We define the horizontal (H) basis by selecting an initial angle of the second analyzer and minimizing coincidences, which corresponds to $90$ degrees \cite{ursin_entanglement-based_2007, pan_experimental_2003}. We set the vertical (V) basis (0 degrees) by maximizing coincidences, while the diagonal (D) and anti-diagonal (A) basis (45, -45 degrees respectively) were set by the waveplate angle at which the coincidence rates for the H and V settings of the other analyzer are equal \cite{ursin_entanglement-based_2007, pan_experimental_2003}. This process results in higher accuracy settings than simply rotating by 45 degrees, due to small inaccuracies in the birefringence of the waveplates \cite{ursin_entanglement-based_2007, pan_experimental_2003}. The results of this measurement are shown in Fig. 
\ref{fig:pol_ent}.  We calculate the visibility as \cite{shalm_strong_2015,shalm_device-independent_2021}

\begin{equation} V=(C_{max}-C_{min})/(C_{max}+C_{min}) .\end{equation} 
where $C_{max}$ and $C_{min}$ are the maximum and minimum coincidences. We assume Poisson statistics when calculating the uncertainty. We measure polarization entanglement visibility in the various bases as D: (99.1 ± 0.1) $\%$, A: (99.1 ± 0.1) $\%$, H: (99.7 ± 0.1) \%, and V: (99.8 ± 0.2) $\%$
The maximum coincidence rate varies depending on the basis primarily due to different system efficiencies for each polarization mode. The high visibility of our polarization state in each basis shows the suitability of our platform for polarization-based protocols.

\subsection{Spectral Correlations}
We measure the joint spectral intensity (JSI) as a first step in characterizing the spectral purity of the signal and idler photons we produce \cite{graffitti_independent_2018,dixon_spectral_2013, chen_efficient_2017, pickston_optimised_2021}. Having spectrally pure single photons is required to obtain high fidelity swapping operations across a quantum network. The joint probability distribution of the signal and idler wavelengths characterizes the spectrum of both photons and their possible correlations \cite{dosseva_shaping_2016, graffitti_independent_2018}. The JSI is separable if the joint spectrum can be represented as the outer product of two independent marginal distributions over the signal and idler \cite{chen_efficient_2017, dosseva_shaping_2016}. The degree to which a JSI is separable, or single-mode, is determined by the Schmidt number $K$ \cite{van_der_meer_optimizing_2020, dixon_spectral_2013, graffitti_independent_2018, pickston_optimised_2021}. The Schmidt number is calculated via the singular values ($\lambda_{i}$) of each two dimensional spectral mode obtained by singular value decomposition \cite{pickston_optimised_2021, chen_efficient_2017}:
\begin{equation}
    K=\sum_{i} \frac{1}{\lambda_{i}^2} \label{eq:K},
\end{equation}
The Schmidt number is $1.0$ for single-mode photons and $>1.0$ for multimode. The purity of a heralded photon in the spectral basis will be $1/K$\cite{chen_efficient_2017, van_der_meer_optimizing_2020}. The joint spectral amplitude (JSA) can have additional phase factors that provide distinguishing information about the photons \cite{graffitti_independent_2018, jeronimo-moreno_control_2009}. These phase correlations primarily arise due to the pump photons having time-dependent spectral properties, such as chirp \cite{jeronimo-moreno_control_2009,sanchez-lozano_relationship_2012}. These temporal correlations can be measured classically through deviations of the pump from the transform limit, through second-order correlation measurements, direct single photon ultra-fast metrology, or through the successive-photon HOM. \cite{bruno_generation_2014, allgaier_highly_2017, kuzucu_joint_2008, davis_pulsed_2017}. 

We measure the JSI using time of flight spectroscopy \cite{davis_pulsed_2017, gerrits_generation_2011}. We herald the JSI photons using the LO. This starts a timing measurement within a window of the 10 ns interpulse time using a fast photodiode. We send each photon through two identical dispersion-compensation modules providing $(1360 \pm 20)$ ps/nm dispersion and 3 dB of loss. We calibrate the relative wavelength using known dispersion, while we determine absolute wavelength by swapping which fiber and dispersion module the signal and idler are sent through. Upon swapping the signal and idler between the two fibers and dispersion modules the spectrum is reflected about the wavelength of the pump divided by two, which is ensured by energy conservation. The results of this timing measurement are shown in Fig. \ref{fig:pol_ent}. We estimate the number of modes in the JSI using singular value decomposition. Using the measured JSI and Eq. \ref{eq:K}, we find a Schmidt number of $1.0089 \pm 0.0002$. This Schmidt number indicates a high degree of spectral purity, indicating these photons can have a high interference visibility when performing swap operations in a network. It is lower than the theoretical number due to imperfections in the pump laser spectrum. The JSI only measures an upper bound on the spectral purity of the photons. To fully characterize the indistinguishability of the photons, HOM interferometry or unheralded second order correlation is required. We rely on HOM interferometry in the main text, but include a second order correlation measurement in the supplement.

% removed g2
\subsection{Hong-Ou-Mandel Interferometry with Successive Photons}
To characterize the indistinguishability of the photons, we perform a successive-photon HOM by interfering heralded signal photons generated in the down-conversion crystal from two different pump pulses separated by approximately 10 ns \cite{graffitti_independent_2018, pickston_optimised_2021, chen_indistinguishable_2019}. This measurement serves as an indicator of the expected swapping fidelity and thus of the suitability of our platform for a photonic quantum repeater. The HOM measurement setup consists of a fixed 10 ns delay arm and a tunable delay arm. We calculate the visibility as shown in Eq. \ref{eq:visT} and uncertainties assuming Poisson counting statistics.
\begin{equation} V = (C_{max}-C_{min})/C_{max} \label{eq:visT} \end{equation} % removed hom setup explanation

The 10 ns delay is the pulse-to-pulse interval of the 100 MHz rep rate laser, allowing two photons from successive pulses to interfere, simulating two independent sources \cite{chen_indistinguishable_2019, graffitti_independent_2018, pickston_optimised_2021}. We herald the idler photons with quasi-number resolution using a beam splitter and two SNSPD detectors. When these two detectors fire 10 ns apart from each other, they herald the injection of two photons separated by 10 ns into the HOM interference setup. An unbalanced Mach-Zehnder interferometer has a $50\%$ chance of delaying each photon by 10 ns. One fourth of the time, the earlier photon takes the long path and the later one takes the short path, such that they arrive at the final beamsplitter simultaneously. Due to the probabilistic nature of the SPDC process, sometimes a single pump pulse produces two or more pairs of photons. If two photons are detected during the time window of the first pulse, no consecutive pair is detected --- the detector dead time is significantly larger than the delay. HOM dip visibility is linearly dependent on the pump power for an experiment using detectors with no number resolution \cite{graffitti_independent_2018,pickston_optimised_2021}. With a number-resolving herald detector, it is possible to post-select on one-photon events, leading to an increase in the visibility. This effect is especially pronounced at higher pump powers. We measure the rate of four-fold coincidences to be $1.85 \pm 0.03 $ four-folds per second at $(48.3 \pm 0.3 )$ mW. The results of this measurement are shown in the left half of Fig. \ref{fig:lo_dip}. The quasi-number resolution affects the dependence of HOM visibility on pump power, but a linear fit is still valid for the parameters of this experiment, which we demonstrate in the supplement.

\begin{comment}
given by Eq. \ref{det2}. Where $\Pi^{2}_{1}$  represents the POVM of a click detector, $ \eta $ is the detector efficiency, and $\mu$ is the per pulse probability measured by a $n \geq 1 $ click detector
\cite{sempere-llagostera_reducing_2022}. We provide this derivation in the supplement. 

\begin{equation}
    V_{\Pi^{2}_{1}} \approx 1-\frac{1}{6} 
    \frac{\left( 3\eta - 4\right)^2}{\eta}
    \mu
     \label{det2}
\end{equation}
\end{comment}
To calculate the visibility in power-dependent measurements, we measure coincidences for a series of delays at high power, tracing out the full HOM dip. Then, for each power, we collect the rate of four-folds at the two delays: one corresponding to the minimum of the dip and the other corresponding to the long-delay baseline level. For each data point in Fig. 3(c), approximately $2000$ counts were collected for the long-delay baseline level and $250$ counts were collected for the dip minimum. The total experiment time of $42$ hours consisted of 3 consecutive trials of a $14$ hour power-dependent visibility measurement. For each pump power, the four-fold coincidences were integrated for a range of approximately $1$ to $6$ hours. We extrapolate a $(96.3 \pm 0.6) \%$ successive-photon HOM dip visibility at zero power without filtering. This infers the visibility expected with an ideal number-resolving herald detector. Having such a high visibility is imperative for maintaining a high purity through successive swapping operations while at a high rate.

\subsection{HOM Interferometry with LO}
\begin{figure}[htbp]
\centering\includegraphics[width=12cm]{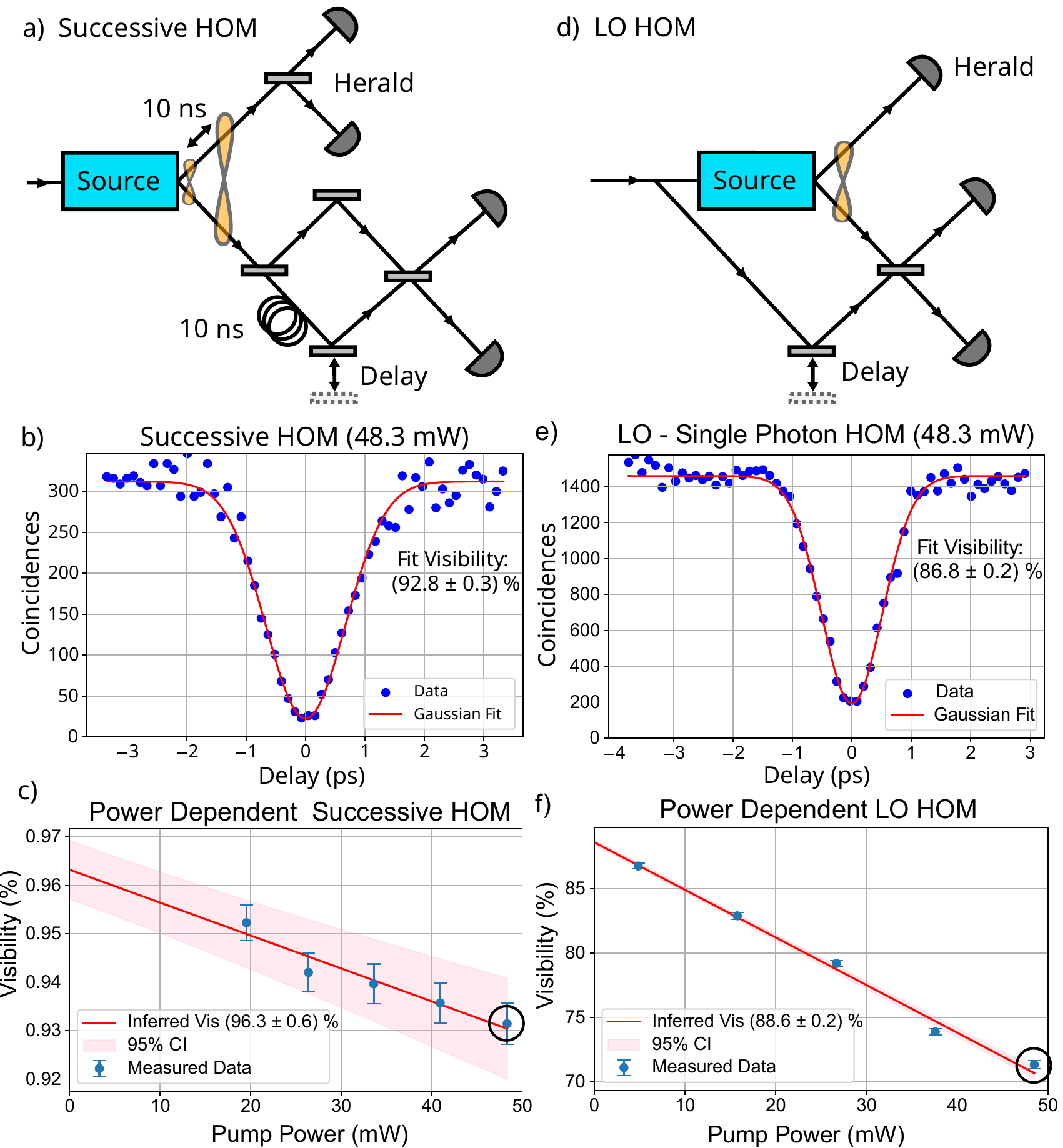}
\caption{The results of our successive-photon and single-photon-LO HOM interference measurements. a) The setup we use to interfere consecutive heralded single photons is shown above. The source emits a pair of photons probabilistically, one of which is measured by two herald detectors. The heralded photons then interfere on a beam splitter. b) A plot of the observed HOM dip as a function of delay for a power of $(48.3 \pm 0.3) $ mW. The coincidences are minimized when the two photons arrive simultaneously at the beam splitter. c) Power dependence of the successive-photon HOM. This allows us to infer the purity of our photons if we had ideal number-resolving detectors. The visibility in this case is limited by multi-photon effects. The four-fold coincidences were integrated for a range of approximately $1$ to $6$ hours as the pump power was reduced. d) A depiction of the heralded single photon and LO interference. e) Measured coincidence rate plotted as a function of time delay between single photon and LO. Similar to the successive-photon case, the LO and the single photon interfere, leading to a bunching effect at the output port. In this case the maximum visibility is limited by multi-photon effects in both the heralded single photon and the LO. f) Measured visibility between single photon and LO plotted as a function of pump power, with LO power held constant. This experiment has a much higher rate due to it being a 3-fold detection event instead of four fold and only corresponded to 1 hour for each power. }
\label{fig:lo_dip}
\end{figure}

Finally, we characterized the quality of our LO by performing a HOM interference between the heralded single photon and the attenuated LO  \cite{bruno_generation_2014, li_experimental_2018 ,xu_characterization_2023}. A photon entering the HOM interferometer is heralded using a detector, while a weak LO is incident every pulse. This experiment provides a proxy measurement for the measurement fidelity of continuous variable measurements. To obtain high measurement fidelities, the LO must identically match the heralded single photon in both spectral and temporal properties. Additionally, if the LO amplitude is high there will be multi-photon events from the LO which will reduce the maximum visibility, as we show in the supplement. We measure the LO to have a rate of $0.0194 \pm .0002$ photons per herald bin through the entire experiment, which will reduce the visibility insignificantly. We measure the three-fold rate to be $(147 \pm 3)$ counts/s at a maximum pump power of $(67 \pm 2)$ mW. The results of this experiment are shown in the right half of Fig. \ref{fig:lo_dip}. The change in visibility as a function of pump power is due to higher-order photon pairs inducing coincidences just as in the successive-photon measurement. We infer a $(88.6 \pm 0.2 )\%$ LO HOM dip visibility at zero power, demonstrating the high degree of indistinguishability between the single photon and the LO it is derived from. This represents to our knowledge the highest HOM interference between a heralded photon and a pulsed LO \cite{bruno_generation_2014, li_experimental_2018 ,xu_characterization_2023}. The inferred zero-power visibility is still limited by the temporal and spectral overlap of the single photon and LO.

\section{Conclusions}

In conclusion, we have developed and demonstrated a high-quality, telecommunication-wavelength platform that is capable of producing indistinguishable heralded single photons, $(96.3 \pm 0.6 )\%$ HOM visibility, high quality polarization entanglement, $(99.11 \pm 0.01) \%$ polarization visibility, or interfere with an LO with $(88.6 \pm 0.2)\%$ visibility. To achieve this, we integrate previous demonstrations of low-loss beam displacer SPDC sources with apodized ppKTP crystals, optimized crystal focusing conditions, and a 1550 nm LO that also acts as a 775 nm pump in a unified platform for optical quantum networking. The high polarization visibility makes this source suitable for protocols that encode entanglement into two degrees of freedom, such as QKD and DI-protocols \cite{yin_entanglement-based_2020, kavuri_traceable_2025, bluhm_single-qubit_2022}. Further, the high successive-photon HOM visibility demonstrates that the heralded single photons are indistinguishable, a key requirement for entanglement swapping protocols \cite{lafler_optimal_2022,pickston_optimised_2021}. Additionally, the high visibility is obtained without filtering allowing us to maintain a high heralded efficiency as detected. Finally, the low-noise stable seed laser used to drive the system can be potentially carrier-phase locked and synchronized with other sources across the network \cite{nardelli_phase-stable_2025}, enabling displacement measurements \cite{paris_displacement_1996} measurements on path-entangled single-photons\cite{alwehaibi_tractable_2025, guo_distributed_2020}.

Unwanted double-pair events limit the HOM visibility at a high rate of single photons. In the future, using a heralding detector with limited single-photon number resolution capabilities will allow double-pair emission events to be filtered out\cite{sempere-llagostera_reducing_2022}. Consequently, the source can be pumped harder and the HOM visibility measured significantly faster while retaining a high purity. Also, higher heralding efficiency can be achieved through low-loss splicing the single-mode fibers together\cite{shalm_strong_2015} instead of butt coupling them. Finally, adaptive filtering can be applied to the local oscillator to better match the spectrum of the single photons, and achieve a higher interference visibility. 

Our source platform is an important step towards creating high-performance, multipurpose, quantum light sources for a variety of telecom-wavelength quantum networking applications\cite{wei_towards_2022, shalm_device-independent_2021,guo_distributed_2020, kavuri_traceable_2025, qi_experimental_2007, gottesman_longer-baseline_2012, alwehaibi_tractable_2025}. A major strength of this approach is the flexibility to implement different qubit encoding and network architectures without needing to make substantial changes to the source, or sacrificing performance. This source is also suited to all-optical quantum repeaters based on path-entanglement due to the potential to lock and stabilize the source lasers across a fiber network\cite{nardelli_phase-stable_2025}. This opens the possibility of carrying out high-performance entangled quantum communication protocols\cite{wei_towards_2022, shalm_device-independent_2021,guo_distributed_2020, kavuri_traceable_2025, qi_experimental_2007, gottesman_longer-baseline_2012, alwehaibi_tractable_2025} across a metropolitan-scale quantum network.

\begin{backmatter}

\bmsection{Acknowledgments}
We would like to thank Paul Kwiat for collaboration on related projects that lead to the development of this source. We would like to thank Cody Mart and Kristen Parzuchowski for helpful discussions. The authors acknowledge support from the AFRL Grand Challenge AWD-22-12-0070, the National Science Foundation QLCI OMA-2016244, the National Science Foundation ECCS 2330228, the University of Colorado Quantum Engineering Initiative, and the National Institute of Standards and Technology. 

\bmsection{Disclosures}
The authors declare no conflicts of interest.

\bmsection{Data Availability Statement}
\bmsection{Data availability} Data underlying the results presented in this paper are not publicly available at this time but may be obtained from the authors upon reasonable request. Simulations can be found at \href{https://app.spdcalc.org/#/?panels=W3sidHlwZSI6ImpvaW50LXNwZWN0cnVtIiwic2V0dGluZ3MiOnsiZW5hYmxlTG9nU2NhbGUiOmZhbHNlLCJoaWdobGlnaHRaZXJvyRZheGlzVMZVV2F2ZWxlbmd0aCIsImF1dG9VcGRhdGUiOnRydWV9fSzJflBhbmVsTG9hZGVyznt9fV0%3D&cfg=eyJhdXRvQ2FsY1RoZXRhIjp0cnVlLMkVUGVyaW9kaWNQb2xpbmfQHkludGVncmF0aW9uTGltaXRzyCFzcGRDb25maWciOnsiY3J5c3RhbCI6IktUUCIsInBtX3R5cGUiOiJUeXBlMl9lX2VvIizIJ1905gCIOTDKE3BoaSI6yxBsZW5ndGgiOjI3NTDLF3RlbXBlcmF0dXJlIjoyxBlvdW50ZXJfcHJvcGFn5QCoIjpmYWxzZSwiZmliZXJfY291cOwA2nB1bXBfd2F2ZchjNzc1xxZiYW5kd2lkxBUwLjbJK2lzdCI6NzDHEHNwZWN0cnVtX3RocmVzaG9sZMQuMDHHH3Bvd2VyIjoxLCJzaWduYWzNazE1NTDJGecBBsoR6AEExw%2FrAI7MTuUAjjHkARXME19wb3Np5gD%2BLTc5MjkuMjU2MTU2NTUwOTU2LCJpZGxlctMqNTcxLjgyMzM4NjUwMjU45AD05wIPX3DlAhBfZW5hYmxlZOkBPsYWxiYiOjQ2LjUyMDMyODUwMDYyMzM0LCJhcG9kaXrlAYrQPcwb5wIgR2F1c3NpYW4izh5md2htIjoxNDbkAOjMGXBhcmHEGs8Wb2lu5AKbW10sImRlZmbFIWnnArtvcuQCo21ldGjkALIiU2ltcHNvxHF0b2xlcmFuY2XlAVAwxGttYXhfZGVw5QF0xhJkZWdyZWUiOjTEDGl2cyI6NTB9fcpeaW9u6gMIbHNfbWlu5AHoNDYuODIsxRFheMQRNTMuxBBp0CFpziFzaXrmAI99fQ%3D%3D} {app.spdcalc.org} 

\end{backmatter}

\bibliography{refs5}

\clearpage
\includepdf[pages=-]{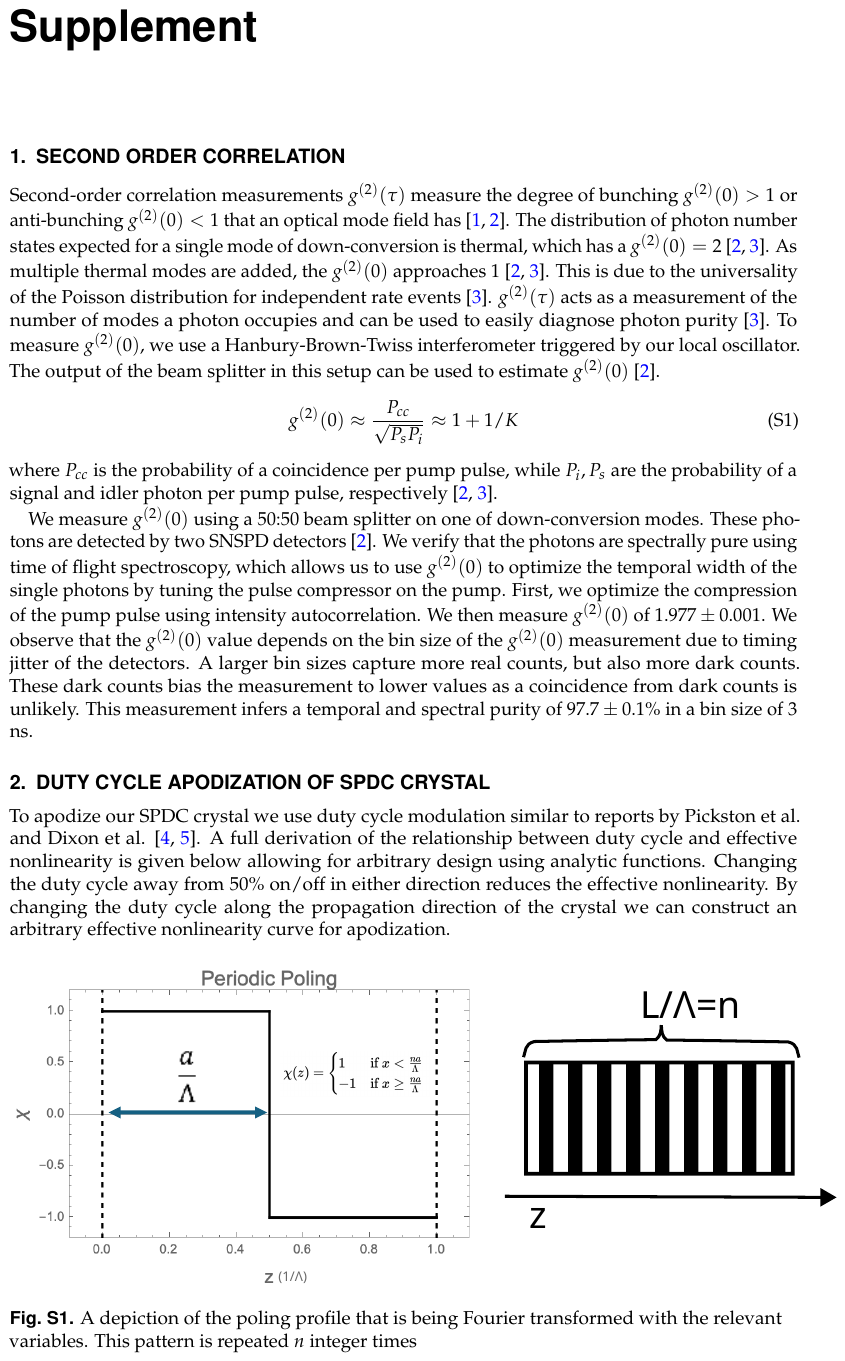}
\end{document}